 \documentclass[preprint,12pt,authoryear]{elsarticle}

\usepackage{amssymb}

\usepackage{multicol}
\usepackage{float}
\usepackage{subcaption}

\usepackage{lipsum}
\makeatletter
\def\ps@pprintTitle{%
 \let\@oddhead\@empty
 \let\@evenhead\@empty
 \def\@oddfoot{}%
 \let\@evenfoot\@oddfoot}
\makeatother

\journal{Nuclear Physics B}

\begin{document}

\begin{frontmatter}



\title{Enhanced navigation systems in GPS denied environments for visually impaired people: A Survey}


\author{Ali Hojjat}
\ead{ahojjat@richmond.ca}
\address{City of Richmond - Local Government}
\address{British Columbia, Canada}

\begin{abstract}
Although outdoor navigation systems are mostly dependent on GPS, indoor systems have to rely upon different techniques for localizing the user, due to unavailability of GPS signals in indoor environments. Over the past decade various indoor navigation systems have been developed. In this paper an overview of some existing indoor navigation systems for visually impaired people are presented and they are compared from different perspectives. The evaluated techniques are ultrasonic systems, RFID-based solutions, computer vision aided navigation systems, ans smartphone-based applications.
\end{abstract}

\begin{keyword}
Indoor navigation \sep Wayfinding \sep Visually impaired navigation 


\end{keyword}

\end{frontmatter}


\section{Introduction}

Moving and travelling in unfamiliar environments is one of the basic and challenging daily activities for visually impaired people  \citep{gharani2017context,Marston2003}. It is challenging because of three major shortcomings due to their impairment, lack of preview, low knowledge of the environment, which could be the same for sighted people as well, and limited access of information for orientation \citep{Golledge97,Harper2003,Helal2001,Marston2003,Ran2004,Roentgen2008}. In order to meet this challenge, substantial number of products, systems, and assistive devices have been developed which could be classified into two major groups: (1) obstacle detection or micro-navigation systems and (2) macro-navigation systems \citep{Katz2012}. Macro-navigation systems generally have four major components: localization, path planning, representation, and interaction \citep{Fallah2013}. As the main purpose of micro-navigation systems is to make BVI users capable to steer away from obstacles, so they are generally not dealing with other components of navigation systems such as localization, path finding, and representation and so on \citep{Katz2012}.

There are numerous ways and technics for helping BVI people to ease their navigation problem. Canes and guide dogs are two common ways to assist BVI users. However, they are used by only a few visually impaired people. Pending the availability of more current information, approximately 109,000 people with vision loss used long canes in 1990 \citep{JVIBnewsservice1994}, while 7,500 individuals used guide dogs in 1995 and 1,500 individuals graduate from a dog-guide user program each year \citep{updates1995demographics}. In addition, the cane has some other shortcomings like no protection against obstacles in upper part of the body, or very limited preview about the ambient environment \citep{Farmer1997}. In other words, both guide dogs and canes are not able to detect overhead objects \citep{cui2010virtual}. A more flexible way is utilizing existing Electronic Travel Aids (ETA's), (e.g. \citep{Bousbia-Salah2011,Ganz2011,Ganz2014,Guerrero2012,Ran2004,Zheng2014}. There are some different types of ETA's that some of them are based on optical triangulation such as LaserCane \citep{benjamin1973laser}. Another type of ETA is acoustic-triangulation-based method, e.g. GuideCane \citep{borenstein2001guidecane}. Minoru 3D webcam \citep{Minoru2010} \fnref{fn1}  ("Minoru 3D webcam," 2010) plus a laptop is ad-hoc ETA based on stereo vision which can recover full depth map. In this paper, I reviewed five recent papers about this topic. The papers functionalities and limitations of some utilized techniques techniques for navigation by visually impaired people are reviewed.

\fntext[fn1]{http://www.minoru3d.com}

\section{BVI Indoor Navigation systems}

\subsection{Ultrasonic-based systems}

\subsection{Computer vision-based system}

\citet{Bousbia-Salah2011} proposed and developed a complete system  to assist visually impaired people for doing their navigational tasks. The system provides the instructions for navigating turn by turn and also could detect obstacles. According to this capabilities, it could be seen that the system tries to integrate both micro and macro navigation systems. In order to make the user aware of the surrounding environment, wide range of sensors are utilized. Using sensors of wearable devices has gain considerable attention in recent years \citep{suffoletto2018using},\citep{gharani2017artificial}. The proposed system has some different parts and it consists of a microcontroller as processor, an accelerometer, a footswitch, a speech synthesizer, a hexadecimal keypad, a mode switch, an ultrasonic cane, two ultrasonic sensors, two vibrators and a power switch. There are some noticeable properties and specifications about the system which make it special. One of them is, it can detect obstacles in a range of 72 degrees in front of the user which seems to be wide enough. Moreover it can detect both static and dynamic obstacles at the distance up to 20 feet. Although detecting both dynamic and static obstacles is a remarkable property, the range of 20 feet could be insufficient. It is also portable, inexpensive, user friendly, and low power consumption.

The system can help people to their wayfinding needs as well. There are two modes in the developed device: record and playback. Playback could be selected in either forward or reverse, which means the system has three possible modes in total. These modes could be selected using the switch. While the system is on record mode, the distance of travelled path by the user is measured and recorded. Besides of that, when user reaches a decision point like an intersection or an entrance door, in which user should take left or right, user press a button and both route and decision stored on the memory. In other word, record mode is learning phase of the system. Instructing the user for helping them to find their way is accomplished in playback mode which is in forward or reverse. The important thing about the reverse mode is the need to change the instruction such as turn left into turn right.

The other major part of the system is obstacle detection subsystem. The obstacle detection has two wearable devices. The first one which should be set up on the shoulder and arms of users, contains two ultrasonic transmitters-receivers and two vibrators. The second part is a cane equipped with ultrasonic sensors and a wheel. The wheel should be in contact with ground and user can perceive the ground type if it is depression, cavity, and the stairs with his hand's tactile sensation intuitively. This system utilizes the ultrasonic sensor in the frequency of 40Hz to transmit pulses, if there would be anything than can block the pulse, which is possibly an obstacle, and returned the pulse back to the receiver, system can measure the time and compute the distance between them. The effective range of this system for detecting obstacles is 0.03 m to 6 m. The results of the computation should be send to the user using vibro-tactile way and speech way for the cane.

The interaction with the user and giving feedback is performed by utilizing vibrator and speech synthesizer. Some information and feedback such as travelled distance, present location and decisions to make is provided to the user by speech synthesizer using a small vocabulary. The feedback about obstacles is returned to the user by vibrotactile. When the system find any obstacle, the vibrotactile is activated and some pulses at a rate which inversely related to the distance to the obstacle is occurred. The information about which side the obstacle is located returned to the user by vibration between the left and the right side. 

PERCEPT \citep{Ganz2011} was proposed as a system to improve the perception of visually impaired people of the indoor environments. The system has some different parts such as passive RFIDs which are installed in the environment, a handheld Android device unit which designed in this project, PERCEPT glove, and PERCEPT server. The Android device is basically the client side of the system and it has some distinct modules such as: Bluetooth module, PERCEPT application, Wi--Fi module, and text to speech engine. The server side of the system is responsible for generating and storing the building information and the RFID tags deployment. The maps are prepared using Quantum GIS. According to the components and parts of the system, it could be classified into three classes and the architecture of the system could be explained as the Environment, the PERCEPT glove and Android client and the PERCEPT server.

Environment has two parts: R-tags and Kiosk. R-tags are passive RFID tags that installed on the door at a specific height (4 feet). They are also used in the kiosks. Kiosks are a sort of data center that user can get useful information about the space. There are a few R-tags in each kiosk that show floor numbers and/or locations (Rooms, restrooms denoted by M and W, emergency exits denoted by X) in the building. 

The second part of the system is PERCEPT Glove and Android Client. It helps user to scan the R-tag. By placing his palm on top of the R-tag, then the glove communicates the chosen destination represented in the R-tag using Bluetooth technology to the Android based Smartphone. There are some buttons on the glove Smartphone that make user able to use the functionalities of the system. The buttons are identifiable by the user taking advantage of different textures. The buttons represent different functionality such as: Help button for the return journey, Replay or Rewind button for repeating the previous instructions in case the user forgets them, instructions button to get next set of instruction.

The third part is PERCEP Server. The spatial data and maps are stored on the server using Postgres database. The shortest path computed and instructions for the navigation prepared for the user. When user touches RFID tags, glove sends data to PERCEPT Server and it retrieves the information about location and the way should be taken to get the destination.

\subsection{Infrared for solving localization problem}

\citet{Guerrero2012} proposed a system for visually impaired people and the main purpose is to provide information to the user in real time to make follow the routes in an indoor environment. The paper tries to answer two questions. First, how to solve localization problem of the user and movement intentions, and second, how to position the obstacles in surrounding area of user. In purpose of addressing these questions, some distinct subsystems are integrated and put together to process the user and environment information. By analyzing the data from the ambient environment, navigational instructions are generated and delivered to the user by messages. Figure 1 shows the main components of the system: an augmented white cane with various embedded infrared lights, two infrared cameras (embedded in a Wiimotes unit), a computer running a software application that coordinates the whole system, and a smartphone that delivers the navigation information to the user through voice messages.

\begin{figure}[H]
\begin{center}

\includegraphics[scale=.5]{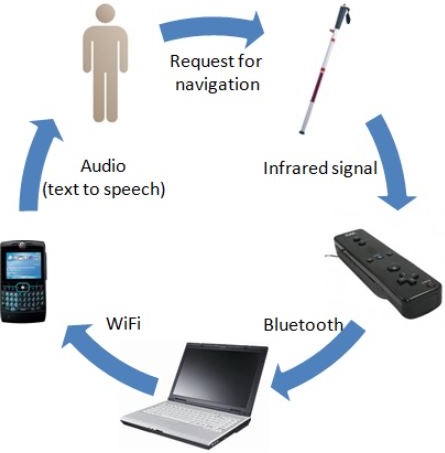}
\caption{Components of the system \citep{Guerrero2012}}

\end{center}
\label{fig1}
\end{figure}

The general idea is the user requests for information by pushing a button on the cane and receives a voice message about navigational instructions on the phone. To clarify, the detailed process is as the button is pushed, the infrared LED's, which are ring of lights around the cane, are activated and they transmit pulses. These signals and pulses are received by Wiimotes. In essence, the Wiimote has infrared camera which can detect the pulses from the cane, then they transmit the received information via Bluetooth to the processing unit which is a laptop. On the words, Wiimotes play a dual role in this system, receiver and transmitter, they can detect the light utmost in 10 meters range. On the laptop machine, the software has spatial data about the environment and the obstacles in the area, so based on the user location and direction system process the data to generate a voice message and send to the user 

In order to handle the navigation process that was mentioned in the previous paragraph, the system used three type of data to provide the information for the user. The first one is the current location of the user. The direction is the second information and system needs to know in which direction the user is moving, and finally the system should have some information about the presence of any possible obstacles in the environment. The paper discuss it in detail how this information is captured and delivered by the system components. Figure 2 shows the process of the data in the system.

\begin{figure}[H]
\begin{center}

\includegraphics[scale=.4]{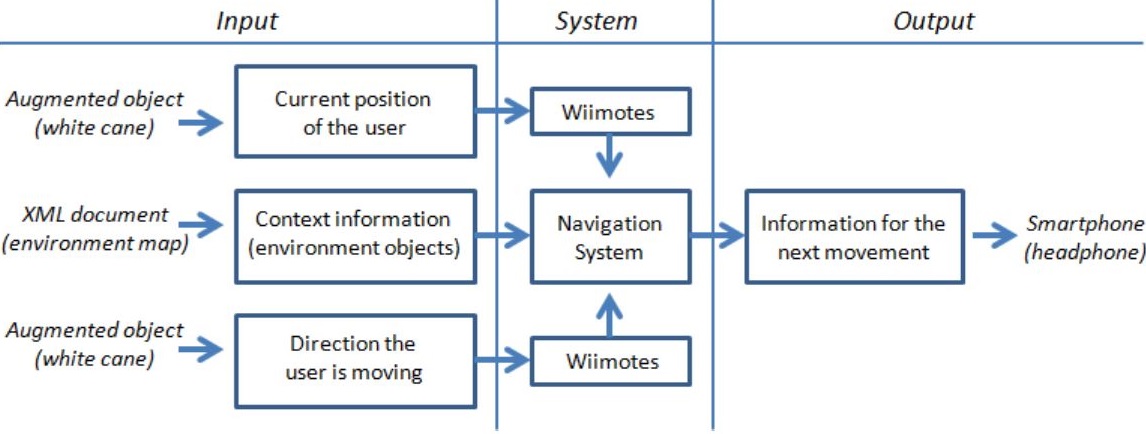}
\caption{Diagram of detecting users position and movements \citep{Guerrero2012}}

\end{center}
\label{fig2}
\end{figure}

The system has a software that can process the XML file containing information about Wiimote position and detected information by them. To be more accurate about the XML file, it should be mentioned it has data about room length and width, relative position of each Wiimote, angle in which the controls were fixed to the walls, and the relative position of each object in the room. The laptop machine can retrieve a map of the room by analyzing the XML file. Thus the environment in which the user is located could be recognized. The position of the user can be computed using triangulation method, it means at least two Wiimotes should cover the user (Figure 3).

\begin{figure}[H]
\begin{center}

\includegraphics[scale=.4]{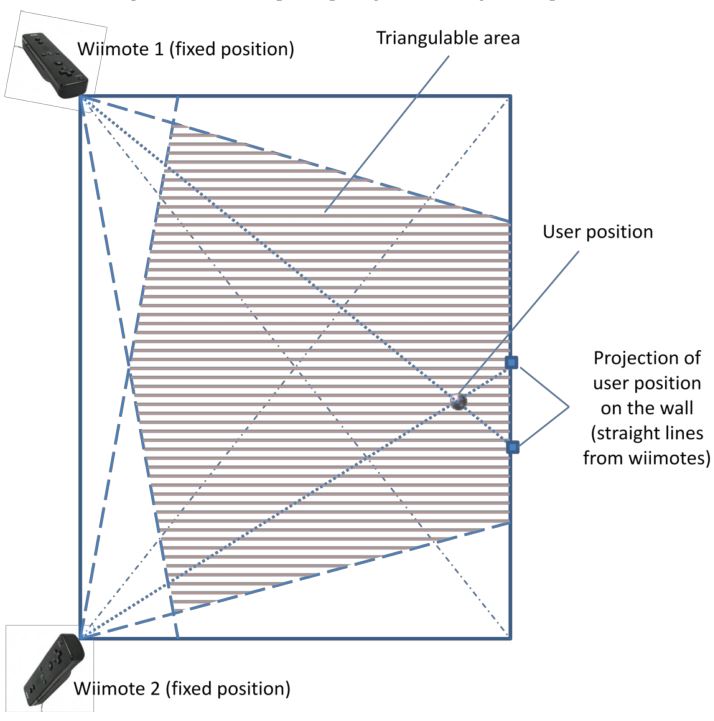}
\caption{Triangulation process \citep{Guerrero2012}}

\end{center}
\label{fig3}
\end{figure}

\subsection{Computer Vision-based Wayfinding Aid}

The problem of evaluating accessibility has been widely studied from different perspectives and in various granularity levels. At geographic scale different models have been developed and implemented \cite{gharani2015enhanced,shi2012spatial,schuurman2010measuring},  In this paper the researchers try to utilize computer vision techniques to solve the challenges of BVI navigation. In essence, computer vision is utilized for the purpose of object detection. There are two types of target objects in this research, the doors and their corresponded signage. Considering the fact that signage plays a critical role in wayfinding the researcher in this paper try to get that sort of information for visually impaired people. The importance of signage and verbal information is due to the importance of distinguishing between similar objects. In other words, users can discriminate between objects in the same class in indoor environments such as elevators, bathrooms, exits, and office doors. They are all doors and geometrically very similar, but for navigation instruction knowing their type is important.

In order to detect the doors, an object detection algorithm was proposed based on general geometric shape in which the configuration of the objects such as edges and corners are processed. After detecting the door, algorithm tries to find the associated text to the door and extract it. The extracted text is then recognized by using off-the-shelf optical character recognition (OCR) software. So, it can help to find out whether the detected door is a bathroom or an office.

The algorithm detects doors in different places, for this purpose and constructing the geometric model of the doors, edges and corners of frames are extracted. The property of these features are having two horizontal lines and two vertical lines between four corners. Canny edge detector \citep{Canny1986} and the corner detector algorithm \citep{ChenHe2008} are utilized. The advantage of this method is avoiding problems such as missing start and end points, sensitivity to parameters, and unwanted merging or splitting of lines. Using perspective geometry is an interesting aspect of the study which helps the user to estimate the relative position of the detected door. The ideal rectangular shape of the doors could be deformed due to the perspective geometry of the camera doors or it could be occluded or just part of them be taken , so there are some assumptions for doing the research first, At least two corners are visible. Second, the vertical lines are both visible and nearly perpendicular to the horizontal axis image frame. The last assumption is about the dimensions of the doors and they should have minimum width and length.

As the doors are detected using the proposed algorithm, the relative position of the door, Left, Front , or Right is returned to the user. Moreover, it is needed to extract the text around the door, convert it to a binary file and use OCR to recognize it. The interesting issue in this paper is about text localization. Topological features from text strings are taken into account. There is an assumption that each text string contains at least three characters, Based on the observation that the distances and the spaces around the characters are consistent like spaces between characters in each word or between words, or between words and signs.

Based on the summarized theory, a wayfinding system is designed. This system contains a camera, a microphone, a portable computer, and a speaker connected by Bluetooth for audio description of objects identified. The camera should be mounted on users' glasses. The user can input the orders into the system and control it by speech input using the microphone. the destination or target pronounced by the user like "find exit door", system process the request and return a speech outputs including "Target found", "Target not found", "Stop function", "Start a new function", etc. if the output is "Target found", the next level of functions includes "Target location", to announce the location and orientation of the target.

\subsection{Smartphone based Indoor Navigation System for the Blind }
\citet{Ganz2014} proposed PERCEPT II  a successor of PERCEPT \citep{Ganz2011} and the substantial improvement is glove is no longer needed and the only needed device to carry is the smartphone. It also added Orientation and Mobility (O\&M) survey tool, which is helpful for giving instruction about the landmarks. Instead and of using Infrared and RFID tags, PERCEPT II deploys Near Field Communication (NFC) tags which is set up on landmarks e.g. doors, elevators, stairs, etc. As the landmarks are detected, some audio feedback is provided to the users. Another amazing improvement is designing and developing a "vision free" interface. This sort of HCI communication is desirable since enables them to interact with the system using built-in accessibility features of the operating system.  There is a server that host the database and user should download the generated instruction from it. The interface of the handheld part is totally vision free and user can use it just by touching it. The information from the landmarks are retrieved when the phone get identification of a landmark using NFC and download the information from the server. The system has an application named O\&M survey tool. It can be executed on an NFC equipped Android tablet. Using this application you can input the landmarks by text, image, sound, and so on. Generating the route instructions and processing the data are its responsibility as well. It is totally vision free.

\section{Comparison of the components of the systems}
One of the important modules of each macro navigation system is localization, i.e. the determination of a user’s position and/or orientation. There are some different techniques for Localization which can be grouped into four different techniques: (1) dead reckoning, (2) direct sensing, (3) triangulation and (4) pattern recognition \citep{Fallah2013}. The first system proposed by \citep{Bousbia-Salah2011} is a micro navigation system. As already mentioned the priority of these systems are detecting obstacles, so there is not explicit localization module, however I believe, as it can retrieve the location from the recorded route, it is a sort of dead reckoning for localization. PERCEPT \citep{Ganz2011}, the second system, uses direct sensing method to find the location. The RFID tags and glove helps user to retrieve the position from the server. \citet{Guerrero2012} in the third article integrated Infrared localization method which is a direct sensing way with triangulation. Computer vision based technique was utilized by \citet{Tian2013} which is part of pattern recognition groups. In PERCEP II, localization is one of those improvements. NFC tags are utilized which changes how the localization process is accomplished, although it is still based on direct sensing, it is another type of localization.

Path planning part is a black box in the last four systems and it seems they rely on some API’s to take the responsibility of that, the first paper does not compute the shortest path, it just record the travelled path and can retrieve and generate the instructions using the stored routes on the system. The other module of the system is interaction. There are two crucial aspects about interaction, first how user can input the request and the second question is, how the user can get feedback from the system. This module is unquestionably a critical part, because with lack of effective communication, the system will be ignored in time, however the other functionalities might be spectacular. \citet{Bousbia-Salah2011} utilized a button on the cane for inserting the request and a switch about the needed mode. The feedback is provided using both speech synthesizer and vibrator together for the users, the handheld device provide a simple keypad to control the switch and select one of the modes. In PERCEPT, user has a glove to scan RFID tags, so it is needed to activate the system to read the tag. PERCEPT II is similar to some extent, although it is technologically more advanced and uses NFC, so user have more options for not holding an extra device. The Android device provide audio messaging to give feedback to the user. PERCEPT has haptic interaction as well. The system developed by  \citet{Tian2013} is able to get speech command for placing the command which is a substantial advantage, and get the feedback by an audio message. Reading signage could improve the communication capabilities.



  \bibliographystyle{elsarticle-harv} 





\end{document}